\documentclass[12pt]{article}
\usepackage{cite}
\usepackage{amsfonts}
\usepackage{amssymb}
\usepackage{amsthm}
\usepackage{amsmath}
\usepackage{pdfsync}
\usepackage{enumerate}
\usepackage{placeins}
\usepackage{lscape}
\usepackage[pdftex,bookmarks,colorlinks,pdfstartview=FitH]{hyperref}
\hypersetup{colorlinks,%
citecolor=blue,%
filecolor=black,%
linkcolor=blue,%
urlcolor=black,%
pdftex}

\title{\textbf{Classification of LRS Bianchi type I Spacetimes via Conformal Ricci Collineations }}
\date{}
\theoremstyle{definition}

\numberwithin{equation}{section}

\author{ \small Tahir Hussain\footnote {Corresponding Author Email: tahirhussain@upesh.edu.pk}, \ Sumaira Saleem Akhtar\ and Fawad Khan\\
\small Department of Mathematics, University of Peshawar, Khyber Pakhtoonkhwa, Pakistan }
\begin{document}
\maketitle
\begin{scriptsize}
\end{scriptsize}
\begin{center}
\noindent\textbf{\large Abstract}
\end{center}
In this paper, we present a complete classification of Locally Rotationally Symmetric (LRS) Bianchi type I spacetimes according to  their Conformal Ricci Collineations (CRCs). When the Ricci tensor is non-degenerate, a general form of the vector field generating CRCs is found, subject to some integrability conditions. Solving the integrability conditions in different cases, it is found that the LRS Bianchi type I spacetimes admit 7, 10, 11 or 15-dimensional Lie algebra of CRCs for the choice of non-degenerate Ricci tensor. Moreover, it is found that these spacetimes admit infinite number of CRCs when the Ricci tensor is degenerate. Some examples of perfect fluid LRS Bianchi type I spacetime metrics are provided admitting non trivial CRCs.\\\\
\begin{footnotesize}
\end{footnotesize}
\section{Introduction}
The main revolutionary idea behind the Einstein's theory of General Relativity (GR) was that the existence of matter induces curvature in a spacetime. This idea coupled geometry and matter through the well known Einstein's Field Equations (EFEs) \cite{[1]}:
\begin{equation}
R_{ab}-\frac{R}{2}\ g_{ab}=G \  T_{ab},  \label{eq:(1.1)}
\end{equation}
where $R_{ab},\ T_{ab}$ and $g_{ab}$ signify Ricci, stress-energy and metric tensors respectively, $R$ is the Ricci scalar and $G$ is the gravitational constant. Finding the exact solutions of EFEs is an important problem in GR which has opened new avenues to our understanding of the universe. The highly non-linear nature of EFEs poses serious problems in finding their exact solutions. Therefore only a reasonable small number of physically interesting exact solutions of EFEs are presented in the literature \cite{[1],[2],[3]}. All these solutions are sought using the assumption that they possess some symmetry. \\
Symmetries (collineations) are vector fields which preserve certain interesting features of spacetime geometry. Some of such features are metric, Riemann, stress-energy and Ricci tensors. For a geometric feature $\Omega,$ to be preserved by a collineation vector field $\xi,$ we require that the Lie derivative of $\Omega$ vanishes in the direction of $\xi$, that is \cite{[4]}:
\begin{equation}
\mathcal{L}_\xi \Omega=0,  \label{eq:(1.2)}
\end{equation}
where $\mathcal{L}_\xi$ signifies Lie derivative operator along the vector field $\xi.$ The $\xi$ is known as a Killing Vector (KV) if $\Omega$ in Eq. (\ref{eq:(1.2)}) denotes a metric tensor $g_{ab}.$ If $\Omega$ in Eq. (\ref{eq:(1.2)}) represents Ricci tensor $R_{ab}$ or Riemann tensor $R^a_{bcd},$ then the corresponding symmetries are called Ricci Collineations (RCs) or Curvature Collineations (CCs) respectively \cite{[5]}. Similarly, one can define Matter Collineations (MCs) by replacing $\Omega$ in Eq. (\ref{eq:(1.2)}) with stress-energy tensor. The KVs are extensively studied to understand the solutions of EFEs and the corresponding conservation laws \cite{[1],[2]}. RCs and CCs are also widely studied in the literature to understand the physics of spacetimes in GR. As a pioneer, katzin \cite{[6]} stated that the CCs are worth investigating as they give additional information about the physics of spacetime that was not given by KVs. An entire and significant work on CCs was published by Hall \cite{[4]} in his book "Symmetries and Curvature Structure in General Relativity", where a classical discussion about other symmetries is also given. As the Ricci tensor is the trace of the Riemann tensor, the RCs have a natural geometric significance \cite{[7],[8],[9]}. The physical importance of RCs is also investigated in the literature. Davis et al. \cite{[10],[11]} discussed this aspect in detail via the conservation laws admitted by particular types of matter fields. They also discussed some interesting applications of the results to relativistic hydrodynamics and plasma physics. Oliver and Davis \cite{[12]} used RCs to obtain conservation expressions for perfect fluids. Some interesting properties of fluid spacetimes are studied by Tsamparlis and Mason \cite{[13]}. Apart from these geometric and physical motivations, RCs are also helpful in the classification of known exact solutions as well as in finding the new exact solutions of EFEs \cite{[14],[15],[16],[17],[18],[19]}. \\
The inheriting symmetries are of particular interest as they provide a deeper insight into the spacetime geometry and are physically significant. These symmetries are used to search the natural relationship between geometry and matter through the EFEs. One of the well known inheriting symmetries is the Conformal Killing Vector (CKV) which satisfies the relation \cite{[4]}:
\begin{equation}
\mathcal{L}_\xi g_{ab}=2\lambda g_{ab},  \label{eq:(1.3)}
\end{equation}
where $\lambda$ is a smooth real valued function of the spacetimes coordinates, called the conformal factor. If $\lambda=0,$ then the solution of Eq. (\ref{eq:(1.3)}) gives KVs and for $\lambda=constant,$ it gives Homothetic Vectors (HVs). A CKV is called a special CKV if $\lambda_{;ab}=0.$ The CKVs
map null geodesics to null geodesics and preserve the structure of null cone. Also, they generate constants of motion along null geodesics for massless particles. This property connects the conformal symmetry with a well defined, physically meaningful conserved quantity. The effects of conformal symmetries can also be seen at kinematic and dynamic levels \cite{[20]}. For a more detailed study of CKVs, we refer \cite{[21],[22],[23],[24],[25],[26]}  \\
After the spacetime metric $g_{ab}$, Ricci tensor $R_{ab}$ is an important candidate as it plays a significant role in understanding the geometric structure of spacetimes. The $\xi$ in Eq. (\ref{eq:(1.3)}) is called a CRC if $g_{ab}$ is replaced by $R_{ab}$. In such a case, Eq. (\ref{eq:(1.3)}) can be written in the explicit form as  \cite{[27],[28]}:
\begin{equation}
R_{ab,c}\ X^c+R_{bc}\ X^c_{,a}+R_{ac}\ X^c_{,b}= 2 \lambda R_{ab}. \label{eq:(1.4)}
\end{equation}
For $\lambda=const.,$ a CRC is called a Homothetic Ricci Collineation (HRC) and if $\lambda=0,$ the CRCs reduce to RCs. If $\xi$ is both a CKV with conformal factor $\alpha$ and a CRC with a conformal factor $\lambda,$ then $\alpha$ and $\lambda$ are related as \cite{[27]}:
\begin{equation}
\alpha_{;ab}=-\lambda \left(R_{ab}-\frac{R}{6} g_{ab}\right),                \label{eq:(1.5)}
\end{equation}
where the term within the braces on right hand side is the well known Schouten tensor, denoted by $A_g.$ The importance of the Schouten tensor in conformal geometry can be viewed in the decomposition of Riemann curvature tensor $R_g$ as $R_g=W_g+A_g\bigodot g,$ where $W_g$ is the conformally invariant Weyl tensor of g and $\bigodot$ is the Kulkarni-Nomizu product. 
The set $CRC(M)$ consisting of all CRCs of a spacetime metric representing a manifold $M$ constitutes a vector space which may be of infinite dimension in case of degenerate Ricci tensor ($det\ R_{ab}=0$). However, for non-degenerate Ricci tensor ($det\ R_{ab}\neq0$), the set $CRC(M)$ forms a finite-dimensional Lie algebra such that $dim \ CRC(M) \leq 15.$ The upper bound of the dimension of $CRC(M)$ is attained if and only if the spacetime is conformally flat \cite{[27]}.\\
Whereas the RCs and HRCs are extensively studied in literature \cite{[29],[30],[31]}, a relatively less interest is taken in the study of CRCs. Only FRW and static spherically symmetric spacetimes are classified according to their CRCs \cite{[17],[27]}. With the hope that classification of spacetimes by conformal collineations of Ricci tensor may be of interest, we provide a complete classification of LRS Bianchi type I spacetimes via CRCs in this paper. The plan of the paper is as follows:\\
In next section, we give a brief introduction of Bianchi type spacetimes and present ten coupled CRC equations for LRS Bianchi type I spacetimes. In section 3, we provide a complete classification of these spacetimes according to their CRCs when Ricci tensor is non-degenerate, while in section 4, the CRC equations are solved for degenerate Ricci tensor. In the last section, we conclude our work with a brief summary and some physical implications of the obtained results.
\section{CRC Equations}
The Bianchi type I, II,...,IX cosmological models are spatially homogeneous spacetimes of dimension $1+3$ admitting a group of motions $G_3$ acting on spacelike hypersurfaces \cite{[1],[32]}. In literature, these spacetimes have been extensively discussed from the symmetry point of view. Out of these spacetimes, the Bianchi type I are the simplest cosmological models for which $G_3$ is the abelian group of translations of the three dimensional Euclidian space. The metric of Bianchi type I spacetimes in the synchronous coordinates has the form \cite{[1]}:
\begin{equation}
ds^2=- dt^2+h^2(t)\ dx^2+k^2(t)\ dy^2+l^2(t)\ dz^2, \label{eq:(2.1)}
\end{equation}
where $h(t),\ k(t)$ and $l(t)$ are functions of time coordinate only. Following are the three basic KVs admitted by the above metric:
\begin{equation}
\xi_{(1)}=\partial_x,\  \xi_{(2)}=\partial_y,\ \xi_{(3)}=\partial_z.    \label{eq:(2.2)}
\end{equation}
If two of the metric functions $h, k$ and $l$ are equal, then the Bianchi type I spacetimes reduce to the important class of LRS spacetimes \cite{[1]} and in such a case an additional rotational KV $z\partial_y-y\partial_z$ is also admitted. For our study, we take $k=l$ in (\ref{eq:(2.1)}) and calculate the CRCs for the resulting LRS Bianchi type I metric. The surviving components of the Ricci tensor for this metric are:
\begin{eqnarray}
R_{00}& = &-\ \frac{h''k+2hk''}{hk}=A(t),  \nonumber \\
R_{11}& = & \frac{hh''k+2hh'k'}{k}=B(t) ,  \nonumber \\
R_{22}& = & R_{33}= \frac{hkk''+h'kk'+hk'^2}{h}=C(t), \label{eq:(2.3)}
\end{eqnarray}
where the primes on the metric functions denote their derivative with respect to $t$. The Ricci curvature scalar $R$ becomes:
\begin{equation}
R=2 \left(\frac{h''}{h}+2 \frac{k''}{k}\right)+2\left(\frac{k'^2}{k^2}+\frac{2h'k'}{hk}\right). \label{eq:(2.4)}
\end{equation}
The Ricci tensor metric for LRS Bianchi type I spacetimes can be expressed as:
\begin{equation}
ds^2_{Ric}=R_{ab} dx^a dx^b=A(t)\ dt^2+B(t)\ dx^2+C(t)\ (dy^2+dz^2), \label{eq:(2.5)}
\end{equation}
which has the original form of LRS Bianchi type I metric. The signature of the above Ricci tensor metric is Lorentzian if $A$ and $B,C$ have opposite signs and it is positive or negative definite if $A, B$ and $C$ have the same signs. Using the EFEs (\ref{eq:(1.1)}) with $G=1,$ we get the following components of stress-energy tensor:

\begin{eqnarray}
T_{00}&=& 2 \frac{h'k'}{hk}+\frac{k'^2 }{k^2}, \nonumber \\
T_{11} &=& -h^2 \biggl( \frac{k'^2}{k^2}+\frac{2k''}{k}\biggr), \nonumber \\
T_{22} &=& T_{33}=-k^2 \biggl(\frac{h''}{h}+\frac{k''}{k}+\frac{h'k'}{hk} \biggr).  \label{eq:(2.6)}
\end{eqnarray}
Using the Ricci tensor components from (\ref{eq:(2.3)}) in Eq. (\ref{eq:(1.4)}), we have the following CRC equations:
\begin{eqnarray}
A' \ \xi^0+2A\  \xi^0_{,0}& = &2\lambda A, \label{eq:(2.7)} \\
A \ \xi^0_{,1}+B\ \xi^1_{,0}& = & 0 ,  \label{eq:(2.8)} \\
A \ \xi^0_{,2}+C\  \xi^2_{,0}& = & 0 , \label{eq:(2.9)} \\
A \ \xi^0_{,3}+C\  \xi^3_{,0}& = & 0 , \label{eq:(2.10)} \\
B'\  \xi^0+2B \ \xi^1_{,1} & = &2\lambda B, \label{eq:(2.11)} \\
B \ \xi^1_{,2}+ C \ \xi^2_{,1} & = & 0 , \label{eq:(2.12)} \\
B \ \xi^1_{,3}+ C \ \xi^3_{,1} & = & 0 , \label{eq:(2.13)} \\
C' \ \xi^0+ 2C \ \xi^2_{,2} & = & 2\lambda C, \label{eq:(2.14)} \\
C\ \left(\xi^2_{,3}+\xi^3_{,2}\right)& = & 0 , \label{eq:(2.15)} \\
C'\  \xi^0+ 2C\  \xi^3_{,3} & = & 2\lambda C. \label{eq:(2.16)}
\end{eqnarray}
The above system contains ten first order coupled partial differential equations which are linear in the conformal vector $\xi=\left(\xi^0,\xi^1,\xi^2,\xi^3\right)$ and the conformal factor $\lambda$. The explicit form of CRCs in LRS Bianchi type I spacetimes can be found by solving this system of equations. In the forthcoming two sections, we solve these equations by considering the non-degenerate and degenerate Ricci tensor cases, without using any form of the stress-energy tensor $T_{ab}.$
\section{CRCs for Non-degenerate Ricci Tensor}
In this section, we consider the Ricci tensor $R_{ab}$ to be non-degenerate, that is $det R_{ab}\neq 0.$ This means that $A,\ B$ and $C$ all are non zero. Using some algebraic calculations, the general solution of Eqs. (\ref{eq:(2.7)})-(\ref{eq:(2.16)}) in terms of some unknown functions of $t$ and $x$ is found as:
\begin{eqnarray}
\xi^0&=&- \frac{C}{A}\ \biggl[\frac{y^2+z^2}{2}\ P^1_t(t,x)+z\ P^2_t(t,x)+y\ P^3_t(t,x) \biggr]+P^0(t,x), \nonumber \\
\xi^1&=&- \frac{C}{B}\ \biggl[\frac{y^2+z^2}{2}\ P^1_x(t,x)+z\ P^2_x(t,x)+y\ P^3_x(t,x) \biggr]+P^4(t,x), \nonumber \\
\xi^2&=& a_2 yz+a_3 \biggl( \frac{z^2-y^2}{2}\biggr)+a_4z+y\ P^1(t,x)+P^3(t,x), \nonumber \\
\xi^3&=& a_2 \biggl( \frac{z^2-y^2}{2}\biggr)-a_3 yz-a_4 y+z\ P^1(t,x)+P^2(t,x), \label{eq:(3.1)}
\end{eqnarray}
with the conformal factor:
\begin{eqnarray}
\lambda &=& -\frac{C'}{2A}\ \biggl[\frac{y^2+z^2}{2}\ P^1_t(t,x)+z\ P^2_t(t,x)+y\ P^3_t(t,x) \biggr]+\frac{C'}{2C}\ P^0(t,x) \nonumber \\
&+& a_2 z-a_3 y+P^1(t,x).  \label{eq:(3.2)}
\end{eqnarray}
The above solution is found subject to the following integrability conditions:
\begin{eqnarray}
\frac{1}{2}\biggl(\frac{C'}{C}-\frac{B'}{B} \biggr)\  P^1_t(t,x)- \frac{A}{B}\ P^1_{xx}(t,x)&=&0 \label{eq:(3.3)} \\
\frac{1}{2}\biggl(\frac{C'}{C}-\frac{B'}{B} \biggr)\  P^2_t(t,x)- \frac{A}{B}\ P^2_{xx}(t,x)-a_2\ \frac{A}{C}&=&0 \label{eq:(3.4)} \\
\frac{1}{2}\biggl(\frac{C'}{C}-\frac{B'}{B} \biggr)\  P^3_t(t,x)- \frac{A}{B}\ P^3_{xx}(t,x)+a_3\ \frac{A}{C}&=&0 \label{eq:(3.5)} \\
2C\ P^1_{tx}(t,x)+B\biggl(\frac{C}{B} \biggr)'\ P^1_x(t,x)&=&0 \label{eq:(3.6)} \\
2C\ P^2_{tx}(t,x)+B\biggl(\frac{C}{B} \biggr)'\ P^2_x(t,x)&=&0 \label{eq:(3.7)} \\
2C\ P^3_{tx}(t,x)+B\biggl(\frac{C}{B} \biggr)'\ P^3_x(t,x)&=&0 \label{eq:(3.8)} \\
\frac{C}{A}\ P^1_{tt}(t,x)+\frac{1}{2}\biggl(\frac{C}{A} \biggr)'\ P^1_t(t,x)&=&0 \label{eq:(3.9)} \\
\frac{C}{A}\ P^2_{tt}(t,x)+\frac{1}{2}\biggl(\frac{C}{A} \biggr)'\ P^2_t(t,x)+a_2&=&0 \label{eq:(3.10)}
\end{eqnarray}
\begin{eqnarray}
\frac{C}{A}\ P^3_{tt}(t,x)+\frac{1}{2}\biggl(\frac{C}{A} \biggr)'\ P^3_t(t,x)-a_3&=&0 \label{eq:(3.11)} \\
\frac{1}{2}\biggl(\frac{B'}{B}-\frac{C'}{C} \biggr)\  P^0(t,x)+P^4_x(t,x)-P^1(t,x) &=&0 \label{eq:(3.12)}\\
\frac{1}{2}\biggl(\frac{A'}{A}-\frac{C'}{C} \biggr)\  P^0(t,x)+P^0_t(t,x)-P^1(t,x) &=&0 \label{eq:(3.13)} \\
A\ P^0_x(t,x)+B\ P^4_t(t,x)&=&0 \label{eq:(3.14)}
\end{eqnarray}
To get the final form of CRCs and the associated conformal factor, we need to solve Eqs. (\ref{eq:(3.3)})-(\ref{eq:(3.14)}). If we denote $\eta_i=\left(\eta_1,\eta_2,\eta_3\right)=\left(\frac{y^2+z^2}{2}, z, y\right)$ and $P^i=\left(P^1,P^2,P^3\right)$, then using the Einstein summation convention we can write the system (\ref{eq:(3.1)}) and the conformal factor given in (\ref{eq:(3.2)}) as follows:
\begin{eqnarray}
\xi^0&=&- \frac{C}{A}\ \eta_i\ P^i_t+P^0, \nonumber \\
\xi^1&=&- \frac{C}{B}\ \eta_i\ P^i_x+P^4, \nonumber \\
\xi^2&=& (\eta_i)_{,2}\ P^i+a_2 yz+a_3 \biggl( \frac{z^2-y^2}{2}\biggr)+a_4z, \nonumber \\
\xi^3&=& (\eta_i)_{,3}\ P^i+a_2 \biggl( \frac{z^2-y^2}{2}\biggr)-a_3 yz-a_4 y, \nonumber \\
\lambda &=& -\frac{C'}{2A}\ \eta_i\ P^i_t+\frac{C'}{2C}\ P^0+P^1+a_2 z-a_3 y. \label{eq:(3.15)}
\end{eqnarray}
Also, the integrability conditions (\ref{eq:(3.3)})-(\ref{eq:(3.14)}) reduce to:
\begin{eqnarray}
\frac{1}{2}\biggl(\frac{C'}{C}-\frac{B'}{B} \biggr)\  P^i_t(t,x)- \frac{A}{B}\ P^i_{xx}(t,x)+k_i\ \frac{A}{C}&=&0, \label{eq:(3.16)} \\
2C\ P^i_{tx}(t,x)+B\biggl(\frac{C}{B} \biggr)'\ P^i_x(t,x)&=&0, \label{eq:(3.17)} \\
\frac{C}{A}\ P^i_{tt}(t,x)+\frac{1}{2}\biggl(\frac{C}{A} \biggr)'\ P^i_t(t,x)-k_i&=&0, \label{eq:(3.18)} \\
\frac{1}{2}\biggl(\frac{B'}{B}-\frac{C'}{C} \biggr)\  P^0(t,x)+P^4_x(t,x)-P^1(t,x) &=&0, \label{eq:(3.19)}\\
\frac{1}{2}\biggl(\frac{A'}{A}-\frac{C'}{C} \biggr)\  P^0(t,x)+P^0_t(t,x)-P^1(t,x) &=&0, \label{eq:(3.20)} \\
A\ P^0_x(t,x)+B\ P^4_t(t,x)&=&0, \label{eq:(3.21)}
\end{eqnarray}
where $k_i=0,-a_2,a_3$ for $i=1,2,3$ respectively. To obtain a complete classification in non-degenerate Ricci tensor case, we solve Eqs. (\ref{eq:(3.16)})-(\ref{eq:(3.21)}) by considering the following cases: \\
\ \ \ \ \ \ \ \  \textbf{(ND1)}\ $A'=B'=C'=0$ \qquad \ \ \ \ \ \ \ \ \ \ \ \ \ \ \ \  \textbf{(ND2)} $A'=C'=0$,\ $B'\neq0$\\
\ \ \ \ \  \ \ \ \textbf{(ND3)}\ $B'=C'=0$,\ $A'\neq0$ \qquad \ \ \ \ \ \ \ \ \ \ \ \ \   \textbf{(ND4)} $A'=B'=0$,\ $C'\neq0$ \\
\ \ \ \ \ \ \ \ \ \textbf{(ND5)}\ $B'=0$,\ $A'\neq0,$\  $C'\neq0$ \qquad \ \ \ \ \ \ \ \ \ \  \textbf{(ND6)} $B'\neq0$,\ $A'=0,$\  $C'\neq0$ \\
\ \ \ \ \ \ \ \ \ \textbf{(ND7)}\ $B'\neq0$,\ $A'\neq0,$\  $C'=0$ \qquad \ \ \ \ \ \ \ \ \ \  \textbf{(ND8)} $A'\neq0$,\ $B'\neq0,$\  $C'\neq0$ \\
As mentioned before, the LRS Bianchi type I spacetimes admit minimum four KVs. Also it is well known that every KV is a RC, so the set of CRCs obtained by solving the system of Eqs. (\ref{eq:(3.16)})-(\ref{eq:(3.21)}) for each of the above cases contain the four basic KVs of these spacetimes. We omit to write the basic algebraic manipulations and summarize the obtained results in Appendix A (Tables 1-8) by presenting the CRCs, the corresponding conformal factor and the constraints on the Ricci tensor components under which LRS Bianchi type I spacetimes admit these CRCs. It is important to mention here that where ever the derivative of the Ricci tensor components $A, B, C$ vanishes, the corresponding Ricci tensor component is assumed to be one. Moreover, the $c_i$'s appearing in all tables denote arbitrary constants.
\section{CRCs for Degenerate Ricci Tensor}
If the Ricci tensor is degenerate, that is $det R_{ab}=ABC^2=0,$ then either $A=B=C=0$ where every direction is a CRC or one of the following possibilities hold.\\
\ \ \ \ \ \ \ \  \textbf{(D1)} $A\neq0$,\ $B\neq0,$\  $C=0$ \qquad \ \ \ \ \ \ \ \ \ \ \ \ \  \textbf{(D2)} $A=0$,\ $B\neq0,$\  $C\neq0$ \\
\ \ \ \ \  \ \ \ \textbf{(D3)}\ $A\neq0$,\ $B=0,$\  $C\neq0$ \qquad \ \ \ \ \ \ \ \ \ \ \ \ \   \textbf{(D4)} $A\neq0$,\ $B=C=0$ \\
\ \ \ \ \ \ \ \ \ \textbf{(D5)}\ $A=C=0$,\ $B\neq0$ \qquad \ \ \ \ \ \ \ \ \ \ \ \ \ \ \ \  \textbf{(D6)} $A=B=0$,\ $C\neq0$ \\
In all the above cases, we directly solve the system of Eqs. (\ref{eq:(2.7)})-(\ref{eq:(2.16)}). In case \textbf{D1}, we are left with the following equations:
\begin{eqnarray}
A' \ \xi^0+2A\  \xi^0_{,0}& = &2\lambda A, \label{eq:(4.1)} \\
A \ \xi^0_{,1}+B\ \xi^1_{,0}& = & 0 ,  \label{eq:(4.2)} \\
\xi^0_{,2} = \xi^0_{,3}=\xi^1_{,2}=\xi^1_{,3}&=&0  , \label{eq:(4.3)} \\
B'\  \xi^0+2B \ \xi^1_{,1} & = &2\lambda B. \label{eq:(4.4)}
\end{eqnarray}
Solving Eqs. (\ref{eq:(4.2)}) and (\ref{eq:(4.3)}), we have
\begin{eqnarray}
\xi^0=- \frac{B}{A}\ f_t(t,x)+g(t),  \ \ \ \ \ \xi^1= f_x(t,x), \label{eq:(4.5)}
\end{eqnarray}
where $f(t,x)$ and $g(t)$ are arbitrary functions of integration. Using the above value of $\xi^0$ in Eq. (\ref{eq:(4.1)}), we get:
\begin{equation}
\lambda=\left(\frac{A'B-2B'A}{2A^2}\right)\ f_t(t,x)-\frac{B}{A}\ f_{tt}(t,x)+\frac{A'}{2A}\ g(t)+g_t(t). \label{eq:(4.6)}
\end{equation}
Simplifying Eq. (\ref{eq:(4.4)}) after using Eqs. (\ref{eq:(4.5)}) and (\ref{eq:(4.6)}), we have:
\begin{equation}
\frac{1}{2} \left(\frac{B}{A}\right)'\ f_t(t,x)+\frac{B}{A}\ f_{tt}(t,x)+f_{xx}(t,x)+\left(\frac{B'}{2B}-\frac{A'}{2A}\right)\ g(t)-g_t(t)=0, \label{eq:(4.7)}
\end{equation}
which is a highly non linear equation and cannot be solved generally. However, one can choose some specific values of the functions $f(t,x)$ and $g(t)$ such that it holds true. With these values of $f(t,x)$ and $g(t),$ we would be able to write the two components $\xi^0$ and $\xi^1$ of CRCs presented in (\ref{eq:(4.5)}) in the final form. The other two components $\xi^2$ and $\xi^3$ are arbitrary functions of $t,x,y$ and $z.$ Hence there are infinite number of CRCs in this case, subject to the differential constraint given in Eq. (\ref{eq:(4.7)}).\\
It is straightforward to solve the CRC Eqs. (\ref{eq:(2.7)})-(\ref{eq:(2.16)}) in the remaining five cases and to see that each case yields infinite number of CRCs. We exclude the basic calculations and present the obtained results of all these cases  in Appendix B (Table 9).
\section{Summary and Discussion}
In this paper, the LRS Bianchi type I spacetimes are completely classified according to their CRCs. It is the extension of our previous work \cite{[29]} about HRCs in the same spacetimes. Setting the conformal factor $\lambda=0,$ we can also obtain the classification of LRS Bianchi type I spacetimes according to their RCs. To achieve a complete classification of LRS Bianchi type I spacetimes according to their CRCs, we have considered both degenerate and non-degenerate Ricci tensor cases. The first outcome of our study is that if the Ricci tensor is degenerate, then the LRS Bianchi type I spacetimes admit infinite number of CRCs. Moreover, our study shows that these spacetimes admit 7, 10, 11 or 15-dimensional Lie algebra of CRCs for the choice of non-degenerate Ricci tensor. In most of the cases considered here, the CRCs are found subject to some differential constraints to be satisfied by the Ricci tensor components. When the number of CRCs is 7, then the CRCs are either reduced to HRCs (see cases \textbf{ND2(b)} and \textbf{ND4(b)}) or one proper CRC is admitted along with one proper HRC and five RCs (see case \textbf{ND4(a)}). In case \textbf{ND7(b)}), we have 10 CRCs, out of which two are proper CRCs, one is a proper HRC and seven are RCs. In case of 11 CRCs, we have four proper CRCs, no proper HRC and seven RCs (see case \textbf{ND4(c)}). In all the remaining non-degenerate cases, we have obtained fifteen CRCs which is the maximum number of CRCs for a 4-dimensional manifold. Setting the conformal factor $\lambda=0$ in all the non-degenerate cases, we can see that the number of independent RCs admitted by LRS Bianchi type I spacetimes is is 4, 5, 6, 7 or 10. In all these cases, the solution of the differential constraints satisfied by the Ricci tensor components would give the exact form of LRS Bianchi type I spacetimes admitting these CRCs. It can be seen that these differential constraints are highly non-linear, so it is not easy to solve them generally. However, our classification shows that CRCs exist in principle. Also, we may choose some specific metric functions $h$ and $k$ so that these differential constraints hold true. As an example, setting $h(t)=t$ and $k(t)=t^2,$ we get the following LRS Bianchi type I spacetime metric:
\begin{equation}
ds^2=-dt^2+t^2\ dx^2+t^4\ \left(dy^2+dz^2\right). \label{eq:(5.1)}
\end{equation}
The above metric satisfies all the constraints of case \textbf{ND5(b)} and hence it admits 15 CRCs as given in Table 4. Out of these 15 CRCs, seven are RCs and remaining eight are proper CRCs. The Ricci scalar $R$ for the above metric is found as $R=\frac{24}{t^2}$ and the components of stress-energy tensor are $T_{00}=\frac{8}{t^2},$ $T_{11}=-8$ and $T_{22}=T_{33}=-4t^2$ which yield a non-degenerate Lorentzian matter tensor metric. One may work on finding such other metrics in the remaining cases.\\
During our classification via CRCs, we have not mentioned any form of the stress-energy tensor $T_{ab}.$ For the physical implications of the obtained results, if we assume a perfect fluid as a source of the stress-energy tensor, then $T_{ab}$ has the form \cite{[14]}:
\begin{equation}
T_{ab}=(p+\rho) u_a u_b+p g_{ab},\label{eq:(5.2)}
\end{equation}
where $p,\rho$ and $u^a$ are the pressure, energy density and the four velocity of the perfect fluid. We choose the fluid velocity as $u^a=\delta^a_0,$ then from Eq. (\ref{eq:(5.2)}), one can obtain $T_{00}=\rho,$ $T_{11}=ph^2$ and $T_{22}=T_{33}=pk^2,$ which give the following perfect fluid matter tensor metric:
\begin{equation}
ds^2_{Perfect}=\rho\ dt^2+p h^2 \ dx^2+p k^2\ (dy^2+dz^2), \label{eq:(5.3)}
\end{equation}
which is positive definite if $\rho>0$ and $p>0.$ Also, the system (\ref{eq:(2.6)}) reduces to:
\begin{eqnarray}
2 \frac{h'k'}{hk}+\frac{k'^2 }{k^2}= \rho,  \ \ \ \   \frac{k'^2}{k^2}+\frac{2k''}{k} = -p,  \ \ \ \  \frac{h''}{h}+\frac{k''}{k}+\frac{h'k'}{hk} = -p.  \label{eq:(5.4)}
\end{eqnarray}
With the help of (\ref{eq:(5.4)}), we can write the Ricci tensor components given in (\ref{eq:(2.3)}) as follows:
\begin{eqnarray}
A(t)=\frac{3p+\rho}{2},   B(t)= \frac{h^2}{2} (\rho-p), C(t)= \frac{k^2}{2} (\rho-p) \label{eq:(5.5)}
\end{eqnarray}
and the Ricci tensor metric, given in (\ref{eq:(2.5)}) becomes:
\begin{equation}
2 ds^2_{Ric}=(3p+\rho)\ dt^2+h^2 (\rho-p)\ dx^2+k^2(\rho-p)\ (dy^2+dz^2). \label{eq:(5.6)}
\end{equation}
The above metric is positive definite if $\rho >0,$ $3p+\rho >0$ and $\rho > |p|$, that is if weak, strong and dominant energy conditions are satisfied. Following is the linear form of a barotropic equation of state $p=p(\rho):$
\begin{equation}
p=(\gamma -1) \rho,\label{eq:(5.7)}
\end{equation}
where $\gamma$ is a constant. The parameters $\gamma=1, 2$ and $\frac{4}{3}$ correspond to dust filled universe, stiff matter and incoherent radiation respectively. Also $\gamma=0$ implies vacuum fluid. For a perfect fluid, the energy conditions are \cite{[14]}:
\begin{eqnarray}
\rho > 0, \ 0\leq p \leq \rho, \  0\leq \frac{dp}{d\rho} \leq 1.\label{eq:(5.8)}
\end{eqnarray}
In our study of CRCs in LRS Bianchi type I spacetimes, perfect fluid is not allowed in some of the degenerate cases, namely \textbf{D1, D3, D5} and \textbf{D6}. Therefore, we may choose some other matter for these cases. In case \textbf{D2}, we have the equation of state $3p+\rho =0$ which violates the strong energy condition. Also, (\ref{eq:(2.3)}) implies $h''k+2hk''=0.$ We may choose some specific functions $h(t)$ and $k(t)$ satisfying this relation to form the exact form of perfect fluid LRS Bianchi type I spacetimes metric. Following is one of such metrics:
\begin{equation}
ds^2=- dt^2+(c_1t+c_2)^2\ dx^2+(c_3t+c_4)^2\ [dy^2+dz^2]; \ c_1\neq0, c_3\neq0.   \label{eq:(5.9)}
\end{equation}
The components of stress-energy tensor for the above metric are $T_{00}=\frac{c_3}{c_3t+c_4} \left(\frac{2c_1}{c_1t+c_2}+\frac{c_3}{c_3t+c_4}\right),$ $T_{11}=-c_3^2\left(\frac{c_1t+c_2}{c_3t+c_4}\right)^2$ and $T_{22}=T_{33}=-c_1c_3 \left(\frac{c_3t+c_4}{c_1t+c_2}\right).$ The Ricci scalar $R$ in this case becomes $R=\frac{2c_3}{c_3t+c_4} \left(\frac{c_3}{c_3t+c_4}+\frac{2c_1}{c_1t+c_2}\right).$ Similarly, the constraints in case \textbf{D4} imply $p=\rho=\frac{A}{2}$ which is the condition for stiff matter. In this case, the system (\ref{eq:(2.3)}) gives $h''k+2h'k'=0$ and $hkk''+h'kk'+hk'^2=0.$ Following is one of the perfect fluid LRS Bianchi type I metric satisfying these equations:
\begin{equation}
ds^2=- dt^2+e^{2t}\ dx^2+e^{-t}\ [dy^2+dz^2].   \label{eq:(5.10)}
\end{equation}
Here the Ricci scalar gets the value $R=\frac{3}{2}.$ The stress-energy tensor components are $T_{00}=-\frac{3}{4},$ $T_{11}=-\frac{3}{4} e^{2t}$ and $T_{22}=T_{33}=-\frac{3}{4} e^{-t}.$ Here the pressure and energy density of the perfect fluid are obtained as $p=\rho=-\frac{3}{4},$ which violate the energy conditions. Also, it is interesting to see that the corresponding matter tensor metric is non-degenerate having negative definite signature.\\
In case \textbf{ND1}, the conditions of perfect fluid imply $3p+\rho=2,$ $\rho=\frac{2}{h^2}+p$ and $h=k.$ Thus the LRS Bianchi type I spacetime metric reduces to the well known conformally flat FRW metric with K=0. A more detailed discussion about RCs and CRCs in FRW metric is given in \cite{[17]}. Similar remarks follow for the case \textbf{ND3} where the perfect fluid conditions yield $h=k$ and $p=\rho-\frac{2}{h^2}.$ In this case, the system (\ref{eq:(2.3)}) gives $A=-\frac{3h''}{h}$ where $h$ satisfies the differential constraint $hh''+2h'^2=1.$\\
In the remaining non degenerate cases, the conditions of a perfect fluid establish relation between pressure $p$ and energy density $\rho$ of the fluid, however the system (\ref{eq:(2.3)}) produces some highly non linear differential constraints involving $h$ and $k$ which cannot be solved easily. One may choose some particular values of $h$ and $k$ satisfying these differential constraints to form the exact form of perfect fluid LRS Bianchi type I spacetime metrics.\\\\
\large {\textbf{Acknowledgments} }\\
\normalsize We are thankful to the unknown referees for their valuable suggestions due to which the manuscript is significantly improved. One of us (S. S. Akhtar) acknowledges the Ph.D indigenous fellowship form Higher Education Commission of Pakistan.
\begin{landscape}
\large {\textbf{Appendix A} }
\begin{table}[h]
\centering
\footnotesize
\begin{center}
\begin{tabular}{|l|c|c|c|}
\hline
 \ \ \ \ Case  \ \ \  & Constraints  &  CRCs \ \ \  & Conformal Factor ($\lambda$)\ \ \ \  \\
\hline
\hline
ND1 & --- & $\xi^0=\frac{c_1}{2}\ \left(t^2-x^2-y^2-z^2\right)+c_2 tx+c_{10}x+c_3t-c_6y-c_8z-a_3ty+a_2tz+c_{11},$ & $a_2z-a_3y+c_1t$ \\
 & & $\xi^1=\frac{c_2}{2}\ \left(x^2-t^2-y^2-z^2\right)+c_1 tx+c_3x-c_{10}t-c_4y-c_7z-a_2xz+a_3xy+c_{12},$ & $+c_2x+c_3$\\
 & & $\xi^2=\frac{a_3}{2}\ \left(t^2+x^2-y^2+z^2\right)+a_2 yz+c_1 ty+c_2 xy+a_4z+c_3y+c_4 x+c_6t+c_5,$ & \\
 & & $\xi^3=\frac{a_2}{2}\ \left(-x^2-t^2-y^2+z^2\right)-a_3 yz+c_1 tz+c_2 xz-a_4y+c_3z+c_7 x+c_8t+c_9.$& \\
\hline
ND2(a) & $B=(c_1t+c_2)^2$& $\xi^0=-c_1 \biggl[ \frac{y^2+z^2}{2} \left(c_{15}\cos c_1x+c_{16}\sin c_1x\right)+y\left(c_{12}\cos c_1x+c_{13}\sin c_1x\right) $& \\
 & & $+z \left(c_9\cos c_1x+c_{10}\sin c_1x\right)+ c_{19}\sin c_1x-c_{20}\cos c_1x \biggr]$&\\
 & & $\frac{\sqrt{B}}{c_1} \left(a_2z-a_3y+c_{17}\right)+\frac{\left(B-c_2^2\right)}{2c_1}\left(c_{15}\cos c_1x+c_{16}\sin c_1x\right),$&\\
& & $\xi^1=\frac{c_1}{\sqrt{B}} \biggl[ \frac{y^2+z^2}{2} \left(c_{15}\sin c_1x-c_{16}\cos c_1x\right)+y\left(c_{12}\sin c_1x-c_{13}\cos c_1x\right) $& $\sqrt{B} \left(c_{15}\cos c_1x+c_{16}\sin c_1x\right)$\\
 & & $+z \left(c_9\sin c_1x-c_{10}\cos c_1x\right)- \left(c_{19}\cos c_1x+c_{20}\sin c_1x\right)$& $+a_2z-a_3y+c_{17}$\\
 & & $+\frac{\left(B+c_2^2\right)}{2c_1^2}\left(c_{15}\sin c_1x-c_{16}\cos c_1x\right) \biggr]+c_{21},$&\\
 & & $\xi^2=a_2yz+\frac{a_3}{2} (z^2-y^2)+a_4z+c_{17}y+y\sqrt{B} \left(c_{15}\cos c_1x+c_{16}\sin c_1x\right) $& \\
 & & $+\sqrt{B} \left(c_{12}\cos c_1x+c_{13}\sin c_1x\right)+\frac{a_3}{2c_1^2} \left(B-c_2^2\right)+c_{14}, $& \\
 & & $\xi^3=-a_3yz+\frac{a_2}{2} (z^2-y^2)-a_4y+c_{17}z+z\sqrt{B} \left(c_{15}\cos c_1x+c_{16}\sin c_1x\right) $& \\
 & & $+\sqrt{B} \left(c_9\cos c_1x+c_{10}\sin c_1x\right)-\frac{a_2}{2c_1^2} \left(B-c_2^2\right)+c_{11}. $& \\
\hline
\end{tabular}
\end{center}
\caption{\footnotesize CRCs for Non-Degenerate Ricci Tensor}
\end{table}
\end{landscape}

\begin{landscape}
\begin{table}[h]
\centering
\footnotesize
\begin{center}
\begin{tabular}{|l|c|c|c|}
\hline
 \ \ \ \ Case  \ \ \  & Constraints  &  CRCs \ \ \  & Conformal Factor ($\lambda$)\ \ \ \  \\
\hline
\hline
ND2(b) & $B=\biggl( c_1t+c_2\biggr)^{2\left(1-\frac{c_3}{c_1}\right)},$      & $\xi^0=c_1t+c_2,\ \ \xi^1=c_3x+c_4,$               & $c_1$ \\
       & where $c_1\neq c_3 \neq 0$                                                 &  $\xi^2=a_4z+c_1y+c_5, \ \ \xi^3=-a_4y+c_1z+c_6.$  &       \\
\hline
ND3 & --- &  $\xi^0=-\frac{1}{\sqrt{A}} \biggl[ \frac{c_7}{2} \left( x^2+y^2+^2\right)+c_{11}y+c_9z+c_{13}x+\left(a_3y-a_2z-c_1x-c_8\right)\int \sqrt{A} dt$&\\
    &     &  $-c_7 \int \left( \sqrt{A} \ \int \sqrt{A} dt \right) dt-c_{15} \bigg],$                                   &\\
    &     &  $\xi^1=\frac{c_1}{2}\ \left(x^2-y^2-z^2\right)-a_3 xy+a_2 xz-c_5y-c_3z+c_8x+\left(c_7x+c_{13}\right)\ \int \sqrt{A} dt$ &$a_2z-a_3y+c_1x$\\
    &     &  $-c_1 \int \left( \sqrt{A} \ \int \sqrt{A} dt \right) dx+c_{14},$                                                    & $+c_7\int \sqrt{A} dt+c_8$ \\
    &     &  $\xi^2=\frac{a_3}{2}\ \left(x^2-y^2+z^2\right)+a_2 yz+c_1 xy+c_8y+a_4z+c_5x+\left(c_7y+c_{11}\right)\ \int \sqrt{A} dt$ & \\
    &     &  $+a_3 \int \left( \sqrt{A} \ \int \sqrt{A} dt \right) dt+c_{12},$                                                       & \\
    &     &  $\xi^3=\frac{a_2}{2}\ \left(-x^2-y^2+z^2\right)-a_3 yz+c_1 xz+c_8z-a_4y+c_3x+\left(c_7z+c_9\right)\ \int \sqrt{A} dt$ & \\
    &     &  $-a_2 \int \left( \sqrt{A} \ \int \sqrt{A} dt \right) dt+c_{10}.$                                                       & \\
\hline
ND4(a) & $C=c_{11}\left(c_5t+c_7\right)^2$  & $\xi^0=c_5xt+c_6t+c_7x+\frac{c_6c_7}{c_5}$,\ \ \ \ $\xi^1=\frac{c_5}{2} (x^2-t^2)+c_6x-c_7t+c_{10}$ & $c_5x+c_6$ \\
 & & $\xi^2=a_4z+c_4,$ \ \ \ \ \ $\xi^3=-a_4y+c_3.$& \\
\hline
ND4(b) & $C=\biggl( c_6t+c_9\biggr)^{2\left(1-\frac{c_1}{c_6}\right)},$ & $\xi^0=c_6t+c_9,\ \ \xi^1=c_6x+c_8,$ & $c_6$ \\
       & where $c_6 \neq 0$ &  $\xi^2=a_4z+c_1y+c_4, \ \ \xi^3=-a_4y+c_1z+c_3.$  &       \\
\hline
\end{tabular}
\end{center}
\caption{\footnotesize CRCs for Non-Degenerate Ricci Tensor}
\end{table}
\end{landscape}

\begin{landscape}
\begin{table}[h]
\centering
\footnotesize
\begin{center}
\begin{tabular}{|l|c|c|c|}
\hline
 \ \ \ \ Case  \ \ \  & Constraints  &  CRCs \ \ \  & Conformal Factor ($\lambda$)\ \ \ \  \\
\hline
\hline
ND4(c) & $C=e^t$  & $\xi^0=2a_3y-2a_2z-ye^{\frac{t}{2}}\ \biggl(c_9\cos\frac{x}{2}+c_{10}\sin\frac{x}{2}\biggr)-ze^{\frac{t}{2}}\ \biggl(c_7\cos\frac{x}{2}+c_8\sin\frac{x}{2}\biggr)-2c_3$& $-\frac{z}{2}\ e^{\frac{t}{2}}\ \biggl(c_7\cos\frac{x}{2}+c_8\sin\frac{x}{2}\biggr)$\\
  & & $\xi^1=e^{\frac{t}{2}}\ \biggl[y\ \left(-c_9\sin\frac{x}{2}+c_{10}\cos\frac{x}{2}\right)+z \left(-c_7\sin\frac{x}{2}+c_8\cos\frac{x}{2}\right) \biggr]+c_{15},$ & $-\frac{y}{2}\ e^{\frac{t}{2}}\ \biggl(c_9\cos\frac{x}{2}+c_{10}\sin\frac{x}{2}\biggr)$\\
  & & $\xi^2=a_2yz+\frac{a_3}{2} \left(z^2-y^2\right)+a_4z+c_3y+2a_3e^{-t}-2e^{-\frac{t}{2}}\ \biggl(c_9\cos\frac{x}{2}+c_{10}\sin\frac{x}{2}\biggr)+c_{11}$& \\
  & & $\xi^3=-a_3yz+\frac{a_2}{2} \left(z^2-y^2\right)-a_4y+c_3z-2a_2e^{-t}-2e^{-\frac{t}{2}}\ \biggl(c_7\cos\frac{x}{2}+c_8\sin\frac{x}{2}\biggr)+c_6$& \\
\hline
ND5(a) & $A=\frac{C'^2}{C^2}$ & $\xi^0=-\frac{C\sqrt{C}}{C'} \biggl[\frac{y^2+z^2}{2}\left(c_1\cos\frac{x}{2}-c_2\sin\frac{x}{2}\right)+y\left(c_9\cos\frac{x}{2}-c_{10}\sin\frac{x}{2}\right)+z \left(c_5\cos\frac{x}{2}-c_6\sin\frac{x}{2}\right) $& $-\frac{\sqrt{C}}{2}\biggl[\frac{y^2+z^2}{2}\left(c_1\cos\frac{x}{2}-c_2\sin\frac{x}{2}\right)$\\
 & & $-\frac{2a_3y}{\sqrt{C}}+\frac{2a_2z}{\sqrt{C}}\biggr]+2\frac{C}{C'} \biggl[ \frac{1}{\sqrt{C}} \left(c_1\cos\frac{x}{2}-c_2\sin\frac{x}{2}\right) -\frac{\sqrt{C}}{2} \left(c_{12}\sin\frac{x}{2}-c_{13}\cos\frac{x}{2}\right)-c_4 \biggr],$& $+y\left(c_9\cos\frac{x}{2}-c_{10}\sin\frac{x}{2}\right)$\\
& & $\xi^1=-\sqrt{C} \biggl[ \frac{y^2+z^2}{2} \left(c_1\sin\frac{x}{2}+c_2\cos\frac{x}{2}\right)+y\left(c_9\sin\frac{x}{2}+c_{10}\cos\frac{x}{2}\right)+z\left(c_5\sin\frac{x}{2}+c_6\cos\frac{x}{2}\right)$& $+z\left(c_5\cos\frac{x}{2}-c_6\sin\frac{x}{2}\right)$\\
 & & $-c_{12}\cos\frac{x}{2}-c_{13}\sin\frac{x}{2} \biggr]-\frac{2}{\sqrt{C}}\left(c_1\sin\frac{x}{2}+c_2\cos\frac{x}{2}\right)+c_{14}, $&$+\left(c_{12}\sin\frac{x}{2}-c_{13}\cos\frac{x}{2}\right)\bigg]$\\
& & $\xi^2=a_2yz+\frac{a_3}{2} \left(z^2-y^2\right)+a_4z+c_4y+\frac{2a_3}{C}-\frac{2y}{\sqrt{C}} \left(c_1\cos\frac{x}{2}-c_2\sin\frac{x}{2}\right)$& $-\frac{1}{\sqrt{C}} \left(c_1\cos\frac{x}{2}-c_2\sin\frac{x}{2}\right)$\\
 & & $ -\frac{2}{\sqrt{C}} \left(c_5\cos\frac{x}{2}-c_6\sin\frac{x}{2}\right)+c_{11},$&\\
 & & $\xi^3=-a_3yz+\frac{a_2}{2} \left(z^2-y^2\right)-a_4y+c_4z-\frac{2a_2}{C}-\frac{2z}{\sqrt{C}} \left(c_1\cos\frac{x}{2}-c_2\sin\frac{x}{2}\right)$&\\
 & & $ -\frac{2}{\sqrt{C}} \left(c_5\cos\frac{x}{2}-c_6\sin\frac{x}{2}\right)+c_8,$&\\
\hline
\end{tabular}
\end{center}
\caption{\footnotesize CRCs for Non-Degenerate Ricci Tensor}
\end{table}
\end{landscape}

\begin{landscape}
\begin{table}[h]
\centering
\footnotesize
\begin{center}
\begin{tabular}{|l|c|c|c|}
\hline
 \ \ \ \ Case  \ \ \  & Constraints  &  CRCs \ \ \  & Conformal Factor ($\lambda$)\ \ \ \  \\
\hline
\hline
ND5(b) & $A=-\frac{C'^2}{C^2}$ & $\xi^0=-\frac{C\sqrt{C}}{C'} \biggl[\frac{y^2+z^2}{2}\left(c_1\cosh\frac{x}{2}+c_2\sinh\frac{x}{2}\right)+y\left(c_9\cosh\frac{x}{2}+c_{10}\sinh\frac{x}{2}\right)+z \left(c_5\cosh\frac{x}{2}+c_6\sinh\frac{x}{2}\right) $& $-\frac{\sqrt{C}}{2}\biggl[\frac{y^2+z^2}{2}\left(c_1\cosh\frac{x}{2}+c_2\sinh\frac{x}{2}\right)$\\
 & & $-\frac{2a_3y}{\sqrt{C}}+\frac{2a_2z}{\sqrt{C}}\biggr]+2\frac{C}{C'} \biggl[ -\frac{1}{\sqrt{C}} \left(c_1\cosh\frac{x}{2}+c_2\sinh\frac{x}{2}\right) +\frac{\sqrt{C}}{2} \left(c_{12}\sinh\frac{x}{2}+c_{13}\cosh\frac{x}{2}\right)-c_4 \biggr],$& $+y\left(c_9\cosh\frac{x}{2}+c_{10}\sinh\frac{x}{2}\right)$\\
& & $\xi^1=-\sqrt{C} \biggl[ \frac{y^2+z^2}{2} \left(c_1\sinh\frac{x}{2}+c_2\cosh\frac{x}{2}\right)+y\left(c_9\sinh\frac{x}{2}+c_{10}\cosh\frac{x}{2}\right)+z\left(c_5\sinh\frac{x}{2}+c_6\cosh\frac{x}{2}\right)$& $+z\left(c_5\cosh\frac{x}{2}+c_6\sinh\frac{x}{2}\right)$\\
 & & $-c_{12}\cosh\frac{x}{2}-c_{13}\sinh\frac{x}{2} \biggr]+\frac{2}{\sqrt{C}}\left(c_1\sinh\frac{x}{2}+c_2\cosh\frac{x}{2}\right)+c_{14}, $&$-\left(c_{12}\sinh\frac{x}{2}+c_{13}\cosh\frac{x}{2}\right)\bigg]$\\
& & $\xi^2=a_2yz+\frac{a_3}{2} \left(z^2-y^2\right)+a_4z+c_4y-\frac{2a_3}{C}+\frac{2y}{\sqrt{C}} \left(c_1\cosh\frac{x}{2}+c_2\sinh\frac{x}{2}\right)$& $+\frac{1}{\sqrt{C}} \left(c_1\cosh\frac{x}{2}+c_2\sinh\frac{x}{2}\right)$\\
 & & $ +\frac{2}{\sqrt{C}} \left(c_9\cosh\frac{x}{2}+c_{10}\sinh\frac{x}{2}\right)+c_{11},$&\\
 & & $\xi^3=-a_3yz+\frac{a_2}{2} \left(z^2-y^2\right)-a_4y+c_4z+\frac{2a_2}{C}+\frac{2z}{\sqrt{C}} \left(c_1\cosh\frac{x}{2}+c_2\sinh\frac{x}{2}\right)$&\\
 & & $ +\frac{2}{\sqrt{C}} \left(c_5\cosh\frac{x}{2}+c_6\sinh\frac{x}{2}\right)+c_8,$&\\
\hline
\hline
\end{tabular}
\end{center}
\caption{\footnotesize CRCs for Non-Degenerate Ricci Tensor}
\end{table}
\end{landscape}

\begin{landscape}
\begin{table}[h]
\centering
\footnotesize
\begin{center}
\begin{tabular}{|l|c|c|c|}
\hline
 \ \ \ \ Case  \ \ \  & Constraints  &  CRCs \ \ \  & Conformal Factor ($\lambda$)\ \ \ \  \\
\hline
\hline
ND6  & $B=C\biggl( \int \frac{1}{\sqrt{C}} dt+\xi\biggr)^2$& $\xi^0=-\sqrt{C} \biggl[ \frac{y^2+z^2}{2} \left(c_{13} \cos x+c_{14} \sin x\right)+y\left(c_7 \cos x+c_8 \sin x\right)$ & $-\frac{C'}{2\sqrt{C}}\biggl[\frac{y^2+z^2}{2} \left(c_{13} \cos x+c_{14} \sin x\right)$\\
 & where $\xi$ is a constant.& $+z\left(c_{10} \cos x+c_{11} \sin x\right)-\frac{B}{2C}\left(c_{13} \cos x+c_{14} \sin x\right)$& $+y\left(c_7 \cos x+c_8 \sin x\right)$\\
 & & $-c_{15} \sin x+c_{16} \cos x+\sqrt{\frac{B}{C}} \left(a_3y-a_2z-c_2\right) \biggr ]$& $+z\left(c_{10} \cos x+c_{11} \sin x\right)$\\
&  & $\xi^1=\sqrt{\frac{C}{B}} \biggl[ \frac{y^2+z^2}{2} \left(c_{13} \sin x-c_{14} \cos x\right)+y\left(c_7 \sin x-c_8 \cos x\right) $ & $+\sqrt{\frac{B}{C}} (a_3y-a_2z-c_2)$\\
 & & $+z\left(c_{10} \sin x-c_{11} \cos x\right)+c_{15} \cos x+c_{16} \sin x \biggr]$ & $-\frac{B}{2C}\left(c_{13} \cos x+c_{14} \sin x\right)$\\
 & & $+\frac{1}{2} \sqrt{\frac{B}{C}} \left(c_{13} \sin x-c_{14} \cos x\right)+c_{17} $& $-\left(c_{15} \sin x-c_{16} \cos x\right)\biggr]$\\
 & & $\xi^2=a_2yz+\frac{a_3}{2} \left(z^2-y^2\right)+a_4z+c_2y+\sqrt{\frac{B}{C}} y\left(c_{13} \cos x+c_{14} \sin x\right) $& $+\sqrt{\frac{B}{C}} \left(c_{13} \cos x+c_{14} \sin x\right)$\\
 & & $\sqrt{\frac{B}{C}} \left(c_7 \cos x+c_8 \sin x\right)+a_3 \int \biggl(\frac{1}{\sqrt{C}} \int \frac{1}{\sqrt{C}} dt \biggr) dt$& $+a_2z-a_3y+c_2$\\
 & & $+a_3 \xi \int \frac{1}{\sqrt{C}} dt+c_9 $&\\
 & & $\xi^3=-a_3yz+\frac{a_2}{2} \left(z^2-y^2\right)-a_4y+c_2z+\sqrt{\frac{B}{C}} z\left(c_{13} \cos x+c_{14} \sin x\right) $& \\
 & & $\sqrt{\frac{B}{C}} \left(c_{10} \cos x+c_{11} \sin x\right)-a_2 \int \biggl(\frac{1}{\sqrt{C}} \int \frac{1}{\sqrt{C}} dt \biggr) dt$&\\
 & & $-a_2 \xi \int \frac{1}{\sqrt{C}} dt+c_{12} $&\\
\hline
\end{tabular}
\end{center}
\caption{\footnotesize CRCs for Non-Degenerate Ricci Tensor}
\end{table}
\end{landscape}

\begin{landscape}
\begin{table}[h]
\centering
\footnotesize
\begin{center}
\begin{tabular}{|l|c|c|c|}
\hline
 \ \ \ \ Case  \ \ \  & Constraints  &  CRCs \ \ \  & Conformal Factor ($\lambda$)\ \ \ \  \\
\hline
\hline
ND7(a) & $A=\frac{B'^2}{B}$&$\xi^0=-\frac{\sqrt{B}}{2B'}\biggl[ \frac{y^2+z^2}{2} \left(c_{11}\cos\frac{x}{2}+c_{12}\sin\frac{x}{2}\right)+y\left(c_9\cos\frac{x}{2}+c_{10}\sin\frac{x}{2}\right)+z \left(c_7\cos\frac{x}{2}+c_8\sin\frac{x}{2}\right) \biggr] $ & $\sqrt{B}\left(c_{11}\cos\frac{x}{2}+c_{12}\sin\frac{x}{2}\right) $\\
 & & $+\frac{B\sqrt{B}}{B'} \left(c_{11}\cos\frac{x}{2}+c_{12}\sin\frac{x}{2}\right)+\frac{\sqrt{B}}{B'}\left(c_{13}\sin\frac{x}{2}-c_{14}\cos\frac{x}{2}\right)+\frac{2B}{B'} \left( a_2z-a_3y+c_2\right),$&$+a_2z-a_3y+c_2$\\
& & $\xi^1=\frac{1}{2\sqrt{B}} \biggl[ \frac{y^2+z^2}{2} \left(c_{11}\sin\frac{x}{2}-c_{12}\cos\frac{x}{2}\right)+y\left(c_9\sin\frac{x}{2}-c_{10}\cos\frac{x}{2}\right)+z \left(c_7\sin\frac{x}{2}-c_8\cos\frac{x}{2}\right) \biggr]$& \\
& &$+\sqrt{B}\left(c_{11}\sin\frac{x}{2}-c_{12}\cos\frac{x}{2}\right)+\frac{1}{\sqrt{B}} \left(c_{13}\cos\frac{x}{2}+c_{14}\sin\frac{x}{2}\right)+c_{15}$& \\
 & & $\xi^2=a_2yz+\frac{a_3}{2} \left(z^2-y^2\right)+a_4z+c_2y+y\sqrt{B}\left(c_{11}\cos\frac{x}{2}+c_{12}\sin\frac{x}{2}\right) $&\\
 & & $+\sqrt{B}\left(c_9\cos\frac{x}{2}+c_{10}\sin\frac{x}{2}\right)+2a_3B+c_6,$&\\
 & & $\xi^3=-a_3yz+\frac{a_2}{2} \left(z^2-y^2\right)-a_4y+c_2z+z\sqrt{B}\left(c_{11}\cos\frac{x}{2}+c_{12}\sin\frac{x}{2}\right) $&\\
 & & $+\sqrt{B}\left(c_7\cos\frac{x}{2}+c_8\sin\frac{x}{2}\right)-2a_2B+c_4$&\\
\hline
ND7(b) & $B=\biggl(a_3\int\sqrt{A}dt+c_1\biggr)^2$ & $\xi^0= \frac{1}{\sqrt{A}} \left(c_7\sin a_3x-c_8\cos a_3x\right)-\frac{1}{a_3} \sqrt{\frac{B}{A}} \left(a_3y-a_2z-c_6\right)$& $a_2z-a_3y+c_6$\\
& & $\xi^1=\frac{1}{\sqrt{B}} \left(c_7\cos a_3x+c_8\sin a_3x\right)+c_9,$& \\
 & & $\xi^2=a_2yz+\frac{a_3}{2} \left(z^2-y^2\right)+a_4z+c_6y+a_3\int \biggl( \sqrt{A} \int \sqrt{A} dt \biggr)dt+c_1\int \sqrt{A} dt+c_2$& \\
 & & $\xi^3=-a_3yz+\frac{a_2}{2} \left(z^2-y^2\right)-a_4y+c_6z-a_2\int \biggl( \sqrt{A} \int \sqrt{A} dt \biggr)dt-\frac{a_2c_1}{a_3}\int \sqrt{A} dt+c_4$& \\
\hline
\end{tabular}
\end{center}
\caption{\footnotesize CRCs for Non-Degenerate Ricci Tensor}
\end{table}
\end{landscape}

\begin{landscape}
\begin{table}[h]
\centering
\footnotesize
\begin{center}
\begin{tabular}{|l|c|c|c|}
\hline
 \ \ \ \ Case  \ \ \  & Constraints  &  CRCs \ \ \  & Conformal Factor ($\lambda$)\ \ \ \  \\
\hline
\hline
ND8(a) &$B=C$ & $\xi^0=-\sqrt{\frac{C}{A}} \biggl[\frac{c_3}{2}(x^2+y^2+z^2)+(a_3y-a_2z-c_1x-c_2) \int \sqrt{\frac{A}{C}} dt$& $-\frac{C'}{2\sqrt{AC}} \biggl[\frac{c_3}{2}(x^2+y^2+z^2)$\\
       &      & $-c_{10}x+c_8y+c_5z-c_3 \int \biggl(\sqrt{\frac{A}{C}} \int \sqrt{\frac{A}{C}} dt \biggr) dt -c_{11} \biggr],$&$+(a_3y-a_2z-c_1x-c_2) \int \sqrt{\frac{A}{C}} dt$\\
       &      & $\xi^1=\frac{c_1}{2}(x^2-y^2-z^2)+c_2x-c_7y-c_4z-a_3xy+a_2xz$& $-c_{10}x+c_8y+c_5z$ \\
       &      & $+c_3x \int \sqrt{\frac{A}{C}} dt-\frac{1}{2c_1} \biggl(c_1 \int \sqrt{\frac{A}{C}} dt+c_{10} \biggr)^2+c_{12},$& $-c_3 \int \biggl(\sqrt{\frac{A}{C}} \int \sqrt{\frac{A}{C}} dt \biggr) dt -c_{11} \biggr]$\\
       &      & $\xi^2=\frac{a_3}{2}(x^2-y^2+z^2)+a_2yz+a_4z+c_1xy+c_7x+c_2y$ & $-a_3y+a_2z+c_1x$\\
       &      & $+(c_3y+c_8)\int \sqrt{\frac{A}{C}} dt+a_3 \int \biggl(\sqrt{\frac{A}{C}} \int \sqrt{\frac{A}{C}} dt \biggr) dt+c_9,$&$+c_3 \int \sqrt{\frac{A}{C}} dt+c_2$\\
       &      & $\xi^3=\frac{a_2}{2}(-x^2-y^2+z^2)-a_3yz-a_4y+c_1xz+c_2z+c_4x$ & \\
       &      & $+(c_1z+c_5)\int \sqrt{\frac{A}{C}} dt-a_2 \int \biggl(\sqrt{\frac{A}{C}} \int \sqrt{\frac{A}{C}} dt \biggr) dt+c_6.$&\\
\hline
ND8(b) &$A=\frac{C^2}{4B} (\frac{B}{C})'$ & $\xi^0=-\frac{2\sqrt{\frac{B}{C}}}{(\frac{B}{C})'} \biggl[\frac{y^2+z^2}{2} (c_8\sin x-c_9\cos x)+y(c_5\sin x-c_6\cos x)+z(c_1\sin x-c_2\cos x)$& $-\frac{C'\sqrt{\frac{B}{C}}}{C(\frac{B}{C})'} \biggl[\frac{y^2+z^2}{2} (c_8\sin x-c_9\cos x)$\\
 & & $-\frac{B}{2C} (c_8\sin x-c_9\cos x)+c_{11}\sin x-c_{12}\cos x+\sqrt{\frac{B}{C}}(a_3y-a_2z-c_{10}) \biggr],$& $+y(c_5\sin x-c_6\cos x)$\\
 & & $\xi^1=-\sqrt{\frac{C}{B}} \biggl[ \frac{y^2+z^2}{2} (c_8\cos x+c_9\sin x)+y(c_5\cos x+c_6\sin x)+z(c_1\cos x+c_2\sin x) $& $+z(c_1\sin x-c_2\cos x)$\\
 & & $+\frac{B}{2C} (c_8\cos x+c_9\sin x)+c_{11}\cos x+c_{12}\sin x \biggr]+c_{13},$& $-\frac{B}{2C}(c_8\sin x--c_9\cos x)$\\
\hline
\end{tabular}
\end{center}
\caption{\footnotesize CRCs for Non-Degenerate Ricci Tensor}
\end{table}
\end{landscape}

\begin{landscape}
\begin{table}[h]
\centering
\footnotesize
\begin{center}
\begin{tabular}{|l|c|c|c|}
\hline
 \ \ \ \ Case  \ \ \  & Constraints  &  CRCs \ \ \  & Conformal Factor ($\lambda$)\ \ \ \  \\
\hline
\hline
 & & $\xi^2=a_2yz+\frac{a_3}{2} (z^2-y^2)+a_4z+c_{10}y+a_3 \frac{B}{2C}$& $+c_{11}\sin x-c_{12}\cos x$\\
 & & $+\sqrt{\frac{B}{C}}y (c_8\sin x-c_9\cos x)+\sqrt{\frac{B}{C}} (c_5\sin x-c_6\cos x)+c_7, $&$+\sqrt{\frac{B}{C}} (a_3y-a_2z-c_{10}) \biggr]$\\
 & & $\xi^3=-a_3yz+\frac{a_2}{2} (z^2-y^2)-a_4y+c_{10}z-a_2 \frac{B}{2C}$& $a_2z-a_3y+c_{10}$\\
 & & $+\sqrt{\frac{B}{C}}z (c_8\sin x-c_9\cos x)+\sqrt{\frac{B}{C}} (c_1\sin x-c_2\cos x)+c_4, $& $+\sqrt{\frac{B}{C}} (c_8\sin x-c_9\cos x)$\\
\hline
ND8(c)& $B=C \biggl(-a_2\int \sqrt{\frac{A}{C}} dt+c_1 \biggr)^2$ & $\xi^0=- \sqrt{\frac{C}{A}} \biggl[ (a_3y-a_2z-c_4) \int \sqrt{\frac{A}{C}} dt$ & $-\frac{C'}{2\sqrt{AC}} \biggl[ (a_3y-a_2z-c_4) \int \sqrt{\frac{A}{C}} dt$\\
 & & $+a_2(c_7 \sin a_2x-c_8 \cos a_2x)-\frac{c_1a_3}{a_2}y+c_1z+\frac{c_1c_4}{a_2} \biggr],$& $+a_2(c_7 \sin a_2x-c_8 \cos a_2x)$\\
 & & $\xi^1=a_2 \sqrt{\frac{C}{B}}\ (c_7 \cos a_2x+c_8 \sin a_2x)+c_9,$& $-\frac{c_1a_3}{a_2}y+c_1z+\frac{c_1c_4}{a_2} \biggr]$\\
 & & $\xi^2=a_2yz+\frac{a_3}{2} (z^2-y^2)+c_4y+a_4z-\frac{c_1a_3}{a_2} \int \sqrt{\frac{A}{C}} dt$& $+a_2z-a_3y+c_4$\\
 & & $+a_3 \int \biggl( \sqrt{\frac{A}{C}} \int \sqrt{\frac{A}{C}} dt\biggr) dt+c_6,$&\\
 & & $\xi^3=-a_3yz+\frac{a_2}{2} (z^2-y^2)+c_4z-a_4y+c_1 \int \sqrt{\frac{A}{C}} dt$&\\
 & & $-a_2 \int \biggl( \sqrt{\frac{A}{C}} \int \sqrt{\frac{A}{C}} dt\biggr) dt+c_2,$&\\
\hline
\end{tabular}
\end{center}
\caption{\footnotesize CRCs for Non-Degenerate Ricci Tensor}
\end{table}
\end{landscape}

\begin{landscape}
\large {\textbf{Appendix B} }
\begin{table}[h]
\centering
\footnotesize
\begin{center}
\begin{tabular}{|l|c|c|c|}
\hline
 \ \ \ \ Case  \ \ \  & Constraints  &  CRCs \ \ \  & Conformal Factor ($\lambda$)\ \ \ \  \\
\hline
\hline
D2(a) & & $\xi^0=\xi^0(x^a),$& \\
 &  $B=C$  & $\xi^1= \frac{c_2}{2} \left( x^2-y^2-z^2\right)-c_4xz-c_5z-c_7xy-c_8y-c_{10}x-c_{11}$  &  $\frac{B'}{2B}\ \xi^0+c_2x-c_4z-c_7y-c_{10}, $  \\
 & & \ \ \ $\xi^2=  \frac{c_7}{2} \left( x^2-y^2+z^2\right)-c_4yz-c_{10}y+c_{12}z+c_2xy+c_8x+c_9, $ &  \\
 & & $\xi^3= \frac{c_4}{2} \left( x^2+y^2-z^2\right)+c_2xz+c_5x-c_{10}z-c_7yz-c_{12}y+c_{14} $&\\
\hline
  D2(b) & $B\neq C$ and $(\frac{B'}{2B}-\frac{C'}{2C})\xi^0+h_x=0$ & $\xi^0=\xi^0(x^a),$ $\xi^1=h,\ \xi^2=c_1\left(c_3z+c_4\right),\ \xi^3=-c_3\left(c_1y+c_2\right)$ &  $\frac{B'}{2B}\ \xi^0+h_x$, where $h=h(x)$   \\
  \hline
  D3 & --- & $\xi^0=- \sqrt{\frac{C}{A}} \biggl[\frac{c_{10}}{2} (y^2+z^2)+c_{12}y+c_{13}z+(c_4y-c_2z) \int \sqrt{\frac{A}{C}} dt $  & $-\frac{C'}{2\sqrt{AC}}\biggl[\frac{c_{10}}{2} (y^2+z^2)+c_{12}y $  \\
& & $- c_{10} \int \left(\sqrt{\frac{A}{C}} \int \sqrt{\frac{A}{C}} dt \right) dt-c_{14} \int \sqrt{\frac{A}{C}} dt-c_{15} \biggr],$ $\xi^1=\xi^1(x^a),$  & $+c_{13}z+(c_4y-c_2z) \int \sqrt{\frac{A}{C}} dt$ \\
  & & $\xi^2=c_{10}y \int \sqrt{\frac{A}{C}} dt+c_4 \int \left(\sqrt{\frac{A}{C}} \int \sqrt{\frac{A}{C}} dt\right) dt +c_{12} \int \sqrt{\frac{A}{C}}dt$ & $- c_{10} \int \left(\sqrt{\frac{A}{C}} \int \sqrt{\frac{A}{C}} dt \right) dt$\\
  & & $+c_{14}y+\frac{c_4}{2} \left(z^2-y^2\right)+c_2 yz+c_5z+c_6,  $& $-c_{14} \int \sqrt{\frac{A}{C}} dt-c_{15} \biggr]$ \\
  & & $\xi^3=c_{10}z \int \sqrt{\frac{A}{C}} dt-c_2 \int \left(\sqrt{\frac{A}{C}} \int \sqrt{\frac{A}{C}} dt\right) dt +c_{13} \int \sqrt{\frac{A}{C}}dt$ & $+c_{10} \int \sqrt{\frac{A}{C}}dt -c_4y$ \\
  & & $+c_{14}z+\frac{c_2}{2} \left(z^2-y^2\right)-c_4 yz-c_5y+c_{15}. $& $+c_2z+c_{14}$\\
 \hline
D4 &    ---    & $\xi^0=h$,\ $\xi^i=\xi^i(x^a); \ i=1,2,3$,  where $h=h(t)$  &  $\frac{A'}{2A}\ h+h_t$  \\
  \hline
  D5 &    ---    & $\xi^1=h$, \ $\xi=\xi^i(x^a); \ i=0,2,3,$ where $h=h(x)$  &  $\frac{B'}{2B}\ \xi^0+h_x$   \\
  \hline
  D6 &    ---    & $\xi^i=\xi^i(x^a);\  i=0,1,$\ $\xi^2=c_1(c_3z+c_4),$\ $\xi^3=-c_3(c_1y+c_2)$ & $\frac{C'}{2C}\ \xi^0$    \\
  \hline
\end{tabular}
\end{center}
\caption{\footnotesize CRCs for Degenerate Ricci Tensor}
\end{table}
\end{landscape}


\begin{thebibliography}{100}
\bibitem{[1]}   H. Stephani, D. Kramer, M. Maccallum, C. Hoenselaers and E. Herlt \emph{Exact Solutions of Einstein's Field Equations} (Second Edition), Cambridge University Press, Cambridge (2003)
\bibitem{[2]}   A. Z. Petrov, \emph{Einstein spaces}, Oxford University Press, Pergamon (1969)
\bibitem{[3]}   C. W. Misnor, K. S. Thorne and J. A. Wheeler \emph{Gravitation}, W.H. Freeman and Company, San Francisco (1973)
\bibitem{[4]} 	G. S. Hall \emph{Symmetries and curvature structure in general relativity}, World Scientific, United Kingdom (2004)
\bibitem{[5]}   A. H. Bokhari and A. R. Kashif \emph{J. Math. Phys.} \textbf{37} 3498 (1996)
\bibitem{[6]}   G. H. katzin and J. Levine \emph{Coloq. Math.} \textbf{26} 21 (1972)
\bibitem{[7]}   I. Yavuz and U. Camci \emph{Gen. Rel. Grav.} \textbf{28} 691 (1996)
\bibitem{[8]}   U. Camci, I. Yavuz, H. Baysal, et al. \emph{Int. J. Mod. Phys. D} \textbf{10} 751 (2001)
\bibitem{[9]}   U. Camci and I. Yavuz, \emph{Int. J. Mod. Phys. D} \textbf{12} 89 (2003)
\bibitem{[10]}  W. R. Davis and G. H. Katzin \emph{Am. J. Phys.} \textbf{30} 750 (1962)
\bibitem{[11]}  W. R. Davis, L. H. Green and L. K. Norris \emph{Nuovo Cimento B} \textbf{34} 256 (1976)
\bibitem{[12]}  D. R. Oliver Jr and W. R. Davis \emph{J. Math. Phys.} \textbf{17} 1790 (1976)
\bibitem{[13]}  M. Tsamparlis and D. P. Mason \emph{J. Math. Phys.} \textbf{31} 1707 (1990)
\bibitem{[14]}  U. Camci and I. T$\ddot{u}$rkyilmaz \emph{Gen. Rel. Grav.} \textbf{36} 2005 (2004)
\bibitem{[15]}  G. Contreras, L. A. Nunez and U. Percoco \emph{Gen. Rel. Grav.} \textbf{32} 285 (2000)
\bibitem{[16]}  I. Yavuz and U. Camci \emph{Gen. Rel. Grav.} \textbf{28} 691 (1996)
\bibitem{[17]}  U. Camci and A. Barnes \emph{Class. Quantum Grav.} \textbf{19} 393 (2002)
\bibitem{[18]}  A. H. Bokhari \emph{Int. J. Theor. Phys.} \textbf{31} 2091 (1992)
\bibitem{[19]}  J. Llosa \emph{J. Math. Phys.} \textbf{54} 072501 (2013)
\bibitem{[20]}  M. Tsamparlis, A. Paliathanasis and L. Karpathopoulos \emph{Gen. Rel. Grav.} \textbf{47} 15 (2015)
\bibitem{[21]}  S. Moopanar and S. D. Maharaj \emph{Int. J. Theor. Phys.} \textbf{49} 1878 (2010)
\bibitem{[22]}  S. Moopanar and S. D. Maharaj \emph{J. Eng. Math.} \textbf{82} 125 (2013)
\bibitem{[23]}  K. L. Duggal and R. Sharma \emph{Nonlinear Analysis} \textbf{63} 447 (2005)
\bibitem{[24]}  R. Maartens, S. D. maharaj and B. O. J Tupper \emph{Class. Quantum Grav.} \textbf{12} 2577 (1995)
\bibitem{[25]}  S. Khan, T. Hussain, A. H. Bokhari and G. A. Khan \emph{Commun. Theor. Phys.} \textbf{65} 315 (2016)
\bibitem{[26]}  S. Khan, T. Hussain, A. H. Bokhari and G. A. Khan \emph{Eur. Phys. J. C} \textbf{75} 523 (2015)
\bibitem{[27]}  U. Camci, A. Qadir and K. Saifullah \emph{Commun. Theor. Phys.} \textbf{49} 1527 (2008)
\bibitem{[28]}  M. Tsamparlis and P.S. Apostolopoulos \emph{Gen. Rel. Grav.} \textbf{36} 47 (2004)
\bibitem{[29]}  T. Hussain, S. S. Akhtar and S. Khan \emph{Eur. Phys. J. Plus} \textbf{130} 44 (2015)
\bibitem{[30]}  T. Hussain, S. S. Akhtar, A. H. Bokhari and S. Khan \emph{Mod. Phys. Lett. A} \textbf{31} 1650102 (2016)
\bibitem{[31]}  T. Hussain, A. Musharaf and S. Khan \emph{Int. J. Geom. Meth. Mod. Phys.} \textbf{13} 1650057 (2016)
\bibitem{[32]}  R. M. Wald, \emph{General Relativity}, Chicago University Press, Chicago (1984)










\end{thebibliography}
\end{document}